\documentclass[usenatbib]{mn2e}

\usepackage{amssymb}
\usepackage{graphicx}

\newcommand{\eqref}{\ref}
\renewcommand{\cite}{\citep}

\def\aaref@jnl#1{{\jnl@style#1}}

\def\aj{\aaref@jnl{AJ}}                   
\def\apj{\aaref@jnl{ApJ}}                 
\def\apjl{\aaref@jnl{ApJ}}                
\def\apjs{\aaref@jnl{ApJS}}               
\def\apss{\aaref@jnl{Ap\&SS}}             
\def\aap{\aaref@jnl{A\&A}}                
\def\aapr{\aaref@jnl{A\&A~Rev.}}          
\def\aaps{\aaref@jnl{A\&AS}}              
\def\mnras{\aaref@jnl{MNRAS}}             
\def\prd{\aaref@jnl{Phys.~Rev.~D}}        
\def\prl{\aaref@jnl{Phys.~Rev.~Lett.}}    
\def\qjras{\aaref@jnl{QJRAS}}             
\def\skytel{\aaref@jnl{S\&T}}             
\def\ssr{\aaref@jnl{Space~Sci.~Rev.}}     
\def\zap{\aaref@jnl{ZAp}}                 
\def\nat{\aaref@jnl{Nature}}              
\def\aplett{\aaref@jnl{Astrophys.~Lett.}} 
\def\apspr{\aaref@jnl{Astrophys.~Space~Phys.~Res.}} 
\def\physrep{\aaref@jnl{Phys.~Rep.}}      
\def\physscr{\aaref@jnl{Phys.~Scr}}       
\def\commat{\aaref@jnl{Comm.~Math.~Phys.}}      
\def\science{\aaref@jnl{Science}}       
\def\cqg{\aaref@jnl{Class.~Quantum Gravity}}        
\def\jpcs{\aaref@jnl{JPCS}}                 
\def\ijmp{\aaref@jnl{Int.~J.~Mod.~Phys.}}           

\title[R-mode in XTE~J1751$-$305]{Implications of an r-mode in  XTE~J1751$-$305: Mass, radius and spin evolution}

\author[Andersson, Jones, \& Ho]{N.\,Andersson$^{1}$\thanks{Email: na@maths.soton.ac.uk}, D. I. Jones$^{1}$ and W. C. G. Ho$^{1}$
\\
$^1$ Mathematical Sciences and STAG Research Centre, University of Southampton,
Southampton SO17 1BJ, UK
}

\date{Accepted 2014 xx. Received 2014 xx;
 in original form 2014 xx}

\begin{document}
\pagerange{\pageref{firstpage}--\pageref{lastpage}} \pubyear{2014}

\maketitle

\label{firstpage}

\begin{abstract}

Recently Strohmayer and Mahmoodifar presented evidence for a coherent oscillation in the X-ray light curve of the accreting millisecond pulsar XTE~J1751$-$305, using data taken by RXTE during the 2002 outburst of this source.  They noted that a possible explanation includes the excitation of a non-radial oscillation mode of the neutron star, either in the form of a g-mode or an r-mode.  The r-mode interpretation has connections with proposed spin-evolution scenarios for systems such as XTE~J1751$-$305.  Here we examine in detail this interesting possible interpretation. Using the ratio of the observed oscillation frequency to the star's spin frequency, we derive an approximate neutron star mass-radius relation which yields reasonable values for the mass over the range of expected stellar radius (as constrained by observations of radius-expansion burst sources).  However, we argue that the large mode amplitude suggested by the Strohmayer and Mahmoodifar analysis would inevitably lead to a large spin-down of the star, inconsistent with its observed spin evolution, regardless of whether the r-mode itself is in a stable or unstable regime.  We therefore conclude that the r-mode interpretation of the observed oscillation is not consistent with our current understanding of neutron star dynamics and must be considered unlikely.  Finally we note that, subject to the availability of a sufficiently accurate timing model, a direct gravitational-wave search may be able to confirm or reject an r-mode interpretation unambiguously, should such an event, with a similar inferred mode amplitude,  recur during the Advanced detector era.

\end{abstract}


\section{Context}

With improved observational sensitivity in both the electromagnetic and gravitational-wave channels, we are approaching the time when we may be able to probe neutron star interiors through asteroseismology. This era may, in fact, already have begun with observations of quasiperiodic oscillations in the tails of magnetar giant flares (see, e.g. \citealt{israel05}, \citealt{SW06}) and the suggestion that the observed oscillations can be mapped to shear modes of the star's crust (e.g. \citealt{SA07}). However, the main lesson learned from this exercise is that the asteroseismology programme is complicated, owing in part to the unknown nature of the state of matter inside the star and the form of the internal magnetic field \citep{toypaper, gabler13}. In order to progress further, we need theoretical models that account for as much of the involved complex physics as possible (or, indeed, palatable because the problem becomes increasingly messy). We also need more observational data. 
In this context, it is interesting to note the recent suggestion by \shortcite{strohmayermahmoodifar13} that a non-radial oscillation mode may have been present in the X-ray data associated with the 2002 discovery outburst of  XTE~J1751$-$305, an accreting neutron star in a low-mass X-ray binary (LMXB) with spin frequency $\nu= 435$~Hz. The observed frequency $0.5727597\times \nu$ is tantalizingly close to the frequency of the star's quadrupole r-mode. In Newtonian gravity, this r-mode would have frequency $2\nu/3$ to leading order of a slow-rotation expansion. Noting this proximity, \shortcite{strohmayermahmoodifar13} discuss whether the observations could be consistent with an r-mode excited in the bulk of the star. This would be an exciting result, given the suggestion that  gravitational-wave emission may drive this particular mode unstable at some critical rotation rate, providing a mechanism to prevent further spin-up by accretion (\citealt{anderssonetal99}; see also \citealt{Ho,haskelletal12}, for a recent assessment of this idea).

At first sight, the r-mode interpretation would seem unlikely, because the observed frequency is close to the mode frequency in a frame co-rotating with the star whereas one might expect a distant observer to see the inertial frame mode frequency, which is $4\nu/3$ to leading order. The first part of any r-mode interpretation has to deal with this issue. The required explanation was provided by \shortcite{NL}, who demonstrated that a nonradial oscillation mode can indeed lead to modulations of an X-ray hotspot being observed at the rotating frame mode frequency. This conclusion is key to the r-mode interpretation of the XTE~J1751$-$305 data. However, this is not sufficient to make the connection credible. 

Strohmayer and Mahmoodifar concluded in part that to interpret the observed oscillation frequency as an r-mode, it was necessary for the star's spin rate to be such that it was close to a resonance between a crustal torsional mode and the r-mode frequency.  Furthermore, for this to be the case, they noted that the star's crust must have a shear modulus approximately twice as large as is generally predicted (cf. the results of \citealt{shearmod}).  One might not be comfortable with such a  sizeable change of a relatively well understood part of the star's physics, but it is a possible explanation. 

However, it turns out that one does not need to tweak crust physics to explain the observed frequency in terms of a global r-mode.  A large correction to the r-mode frequency is caused by the combined relativistic effects of the gravitational redshift and rotational frame-dragging. These effects were calculated some time ago by \citet{LFA03}. According to their analytic expression for a uniform density barotropic model (equation~36 of \citealt{LFA03}) the quadrupole ($m=2$) r-mode should have a co-rotating r-mode frequency $\kappa\times\nu$  where
\begin{equation}
\kappa = {2 \over 3} \left[ 1 - {8\over 15} \left( {M\over R}\right) \right] ,
\label{pNfreq}\end{equation}
to first post-Newtonian (1PN) order.  This shows that, for a typical neutron star compactness of $M/R\approx 0.2$, the relativistic corrections is at the 10\% level. This is significantly larger than the rotational and crust corrections (unless one invokes resonances) considered in \citet{strohmayermahmoodifar13}. Moreover, the relativistic effects tend to \emph{lower} the (rotating frame) mode frequency, exactly as required to explain the observations, which require $\kappa \approx 0.57$. This motivates us to take a closer look at the problem. 

In Section~\ref{mass} we go beyond the rough estimate of equation (\eqref{pNfreq})  and argue that we can use the observational results to constrain  the relationship between mass and radius for this neutron star.  If one additionally assumes that the radius lies in the range inferred from radius-expansion burst sources \citep{LS}, estimates for the mass itself are obtained.
We show  that this procedure yields a sensible result.  
However, this is only part of the story---it is also necessary to explain the observed variation in spin frequency for this system.  In Section \ref{observations} we first consider the issue of spin evolution including only standard accretion torques and magnetic fields, without any r-mode excitation.  We show that the standard models fail to explain the observed frequency variation of XTE~J1751$-$305, as different estimates/constraints on the star's magnetic field fail to agree; we find the same disagreement for another LMXB, IGR~J00291+5934.  In Section \ref{scenarios} we consider the effect of adding an r-mode into the model, with the large amplitude indicated by the Strohmayer \& Mahmoodifar analysis.  We show that, in all plausible scenarios (i.e., stable and unstable and unsaturated and saturated), the inclusion of the r-mode over the duration of the X-ray outburst (10 days or so) in fact makes the observed spin variation much \emph{more} difficult to explain (Strohmayer \& Mahmoodifar noted this problem while considering only the unstable scenario).  We therefore conclude that the r-mode interpretation of the data is hard to sustain, but note that future gravitational wave detectors could, if a sufficiently accurate timing model is available, be able to directly test if an r-mode is present at a similar level in any future outburst of XTE~J1751$-$305 or a similar system.

\section{Inferring the star's mass}
\label{mass}

\begin{figure}
\begin{center}
\includegraphics[scale=0.42,clip]{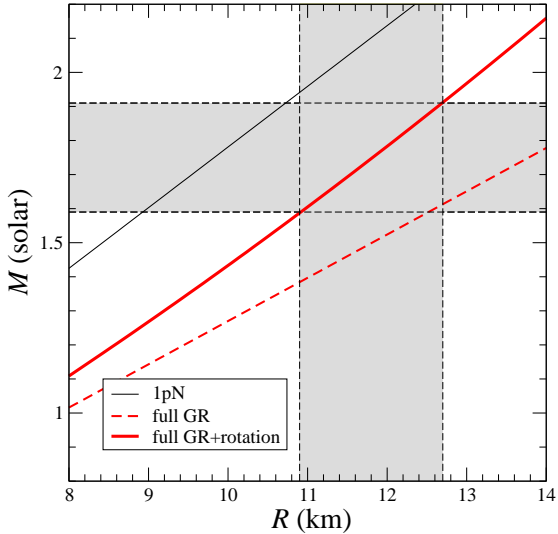}
\caption{Various relations between mass and radius obtained from the observed ratio of spin frequency to `mode' frequency, assuming the mode to be the $l=m=2$ r-mode.  The solid black line includes the effects of General Relativity at the first post-Newtonian level, but neglects the effects of rapid rotation.  The dashed red line includes the effects of General Relativity but again neglects rapid rotation.  The solid red line includes both General Relativity and an (Newtonian-based) estimate of the effects of rapid rotation.  These curves simply give our estimates of the  mass as a function of radius---no attempt has been made to bracket uncertainties; see text for discussion. }
\label{mrplane}
\end{center}
\end{figure}

The ratio of the  $l=m=2$ (rotating frame) r-mode frequency to spin frequency is given by $\kappa = 2/3$ only in the highly idealised case of small oscillations of a slowly rotating perfect fluid star, modelled using Newtonian gravity.  More realistic models will lead to different $\kappa$-values, allowing us to use the observed $\kappa \approx 0.573$ value to learn something of the physics of the real system.  As discussed above, the effect of General Relativity is significant.  The first post-Newtonian (1PN) result given in equation (\ref{pNfreq}), strictly valid only for a uniform density star (i.e.\ a $n=0$ polytrope), leads to an estimate for the compactness $M/R \approx 0.264$.  This can be viewed as a constraint in the mass-radius plane, as represented by the black line in Figure \ref{mrplane}.  

However, perusal of Figure 1 of \citet{LFA03}, which gives the mode frequency as a function of compactness for a \emph{fully relativistic} (numerical) calculation,  shows a significant \emph{non-linear} variation of $\kappa$ with compactness, indicating that the 1PN result is \emph{not} very accurate for stars with realistic compactness values.  Using instead the full relativistic  result, easily read-off from Figure 1 of \citet{LFA03}, we obtain the (presumably) more accurate estimate  $M/R \approx 0.187$.   This corresponds to the dashed red curve of Figure \ref{mrplane}.  This constraint is strictly only valid for uniform density stars, but as shown in Figure 2 of \citet{LFA03}, $\kappa$ varies by only a few percent when the polytropic index $n$ is varied (at fixed $M/R$) over the interval $0 < n < 1.5$, so it seems the variation of $\kappa$ with $M/R$ is much more significant than its variation with the equation of state.

As described in \citet{LMO99}, there will be rotational corrections to $\kappa$ too, of order 
\begin{equation}
\label{eq:Omega_squared_over_rho}
\frac{\Omega^2}{\pi G \bar\rho_0} = 5.33 \times 10^{-2} \, \left(\frac{\nu}{435 \, \rm Hz}\right)^2
\left(\frac{R}{10^6 \, \rm cm}\right)^3 \left(\frac{1.4 M_\odot}{M}\right) ,
\end{equation}
where the angular frequency is $\Omega = 2\pi \nu$ and
$\bar\rho_0$ is the average density of the corresponding non-rotating star of mass $M$ and radius $R$.  This indicates that for a star spinning as rapidly as XTE~J1751$-$305, rotational corrections are not dominant, but not negligible either.   To account for the effect of rotation into our analysis, we write the actual $\kappa$ value as the sum of its Newtonian part and the relativistic and rotational corrections:
\begin{equation}
\kappa = \kappa_{\rm N} + \delta \kappa_{\rm GR} + \delta \kappa_{\rm rot} .
\end{equation}
The combination $\kappa_{\rm N} + \delta \kappa_{\rm GR} \equiv \kappa_{\rm GR}$, as a function of $M/R$, can be read-off from Figure 1 of \citet{LFA03}.  The rotational corrections are estimated by \citet{LMO99}, who wrote their result in the form
\begin{equation}
\delta \kappa_{\rm rot} = \kappa_2 \frac{4\Omega^2}{3G} \frac{R^3}{M} .
\end{equation}
In \citet{LMO99} it was found that, for stars with $M = 1.4 M_\odot$ and for a variety of realistic equations of state, $0.26 \lesssim \kappa_2 \lesssim 0.32$, so we take $\kappa_2 \approx 0.29$ as a representative value.  
We then have
\begin{equation}
\label{eq:kappa_full}
\kappa = \kappa_{\rm GR}(M/R) +  \kappa_2 \frac{4\Omega^2}{3G} \frac{R^3}{M} .
\end{equation}
Using the observed value $\kappa \approx 0.573$, the numerical calculation of $\kappa_{\rm GR}$ of Figure 1 of \citet{LFA03}, and the value $\kappa_2 = 0.29$, the value of $R$ can then be varied over some sensible range, and equation~(\ref{eq:kappa_full}) is solved numerically for $M(R)$.   The results of such a calculation are depicted by the solid red line in Figure \ref{mrplane}.  Note that our procedure is  inconsistent, in that we use the $n=0$ results of \citet{LFA03} for $\kappa_{\rm GR}$, but an average over realistic equations of state for a $1.4 M_\odot$ star from the results of \citet{LMO99} for $\kappa_2$.  However, as noted above, in both cases the variation in $\kappa$ with respect to variations in the equation of state is small, at the level of a few percent, so errors introduced by this inconsistency are presumably at a similar level, much smaller than the variation in $M$ when $R$ is varied over an astrophysically-plausible interval.  
Given the relatively crude nature of our approximations, a detailed error analysis is clearly not appropriate, but given the several percent variation of $\kappa_{\rm GR}$ of the relativistic correction with polytropic index, and the variation in the rotational parameter $\kappa_2$ over realistic equations of states,  the error in this $M(R)$ curve due to our lack of knowledge of the real equation of state is  plausibly smaller than 10\%. 

Assuming that the inferred compactness reflects the nature of the neutron star, we can combine the result with the radius constraint from X-ray burst sources. According to the most recent analysis by \citet{LS}, the allowed radius range for all neutron stars between $1.2 M_\odot$ and $2.0 M_\odot$ is 10.9 to 12.7~km\footnote{Note that a lower radius range is found by \citet{guillotetal13}.  However an important systematic effect (as well as others) is the assumption of a hydrogen atmosphere when fitting X-ray data.  A helium atmosphere leads to a larger inferred radius (see, e.g. \citealt{servillatetal12,catuneanuetal13}).  This difference in composition is approximately taken into account in the analysis of \citet{LS}.}.
In other words, one would expect 
$R= 11.8\pm0.9$~km, or, more appropriately given our likely level of accuracy, $R= 12\pm1$~km. Combining this result with the inferred compactness, we conclude that the mass of the XTE~J1751$-$305 neutron star should lie in the range $1.59-1.91 M_\odot$, with a preferred value of 
$M \approx 1.8 M_\odot$. This is a perfectly reasonable mass for a neutron star that has accreted enough mass to be spun up to the observed fast spin.  More detailed calculations, especially concerning the fast rotation correction in General Relativity (perhaps building on \citealt{gaertig}) are needed to improve on this.

The close proximity (in frequency) of a crustal toroidal mode could have a significant effect on the r-mode frequency, and change the above estimates.  However, as  Strohmayer \& Mahmoodifar noted, proximity to such an avoided crossing would require an unexpectedly large value for the crustal shear modulus, so this solution does not seem very likely.  There are other pieces of physics, beyond the inclusion of full relativity, rapid rotation, and the crust, that could affect the r-mode frequency.  These include magnetic fields and superfluidity.  The weak magnetic fields of LMXBs indicate that magnetic corrections should be negligible [see, e.g. \citealt{letal10}, for normal (rather than superconducting) interiors at least].  Similarly, studies of Newtonian stars with superfluid cores indicate that the r-mode we are considering here is relatively insensitive to this aspect \cite{passamontiandersson12}.

\section{Observations and spin evolution}  
 \label{observations}

Having established that the observed frequency would be consistent with a global r-mode oscillation in 
XTE~J1751$-$305, let us now turn to the implications for and constraints from the measured spin evolution of this system.  We will begin by applying `standard' accretion theory to this system, which involves only accretion and magnetic fields, deferring the inclusion of r-modes to Section \ref{scenarios}. 

The 2002 X-ray outburst in XTE~J1751$-$305 occurred from April 3 to April 30, had a rise time of less
than 4~d, peaked around April 4--5, and declined exponentially with a
decay timescale of $\approx 7\mbox{ d}$ \cite{markwardtetal02}.
Pulsations at the spin-frequency of 435~Hz were only detectable until April 14,
with the pulse fraction decreasing from 5\% to 3\% during this early period,
and no pulsations were detected afterwards, including during a small outburst
that occurred April 28--30 (pulse fraction limit $<5.5\%$).
A reanalysis by \citet{papittoetal08} of the outburst found that
coherent pulsations could be detected for 9~d (after which the flux decreased
dramatically) and that a frequency increase (spin-up) occurred during this time. They estimated the spin-up rate to be 
\begin{equation}
\dot{\nu}_{\mathrm{su}}=(3.7\pm 1.0)\times 10^{-13}\mbox{ Hz s$^{-1}$},
 \label{eq:suxte}
\end{equation}
with an error  at the 90\% confidence level. The total change in the spin-frequency was found to be 
\begin{equation}
\Delta\nu=2.8\times 10^{-7}\mbox{ Hz}.
 \label{eq:Delta_nu_observed}
\end{equation}
The analysis in \citet{papittoetal08} was not able to establish whether
there was a dependence of $\dot{\nu}$ on the mass accretion rate $\dot{M}$.
We note that the distance to XTE~J1751$-$305 is estimated to be
$\sim 6-9\mbox{ kpc}$ \citep{markwardtetal02,papittoetal08}, which yields
a peak 2-10~keV luminosity of $<1.3\times 10^{37}\mbox{ erg s$^{-1}$}$ for
the 2002 outburst \citep{riggioetal11},
corresponding to $\dot{M}<1.1\times 10^{-9}M_\odot\mbox{ yr$^{-1}$}$.

XTE~J1751$-$305 has subsequently exhibited outbursts in 2005, 2007, and 2009.
These outbursts were weaker (at 14\%, 18\%, and 27\%, respectively,
of the 2002 flux) and shorter (with decay timescales $\approx 2\mbox{ d}$)
\cite{riggioetal11}.
Pulsations were only detected in the 2009 outburst for 1~d, with an amplitude
of 7--8\%, and there was no significant frequency evolution within this
outburst.
The analysis in \citet{riggioetal11} suggests a long-term decrease in frequency (spin-down)
of the pulsations detected in 2002 and those in 2009, with
\begin{equation}
\dot{\nu}_{\mathrm{sd}}=-(5.5\pm 2.0)\times 10^{-15}\mbox{ Hz s$^{-1}$}.
 \label{eq:sdxte}
\end{equation}
The picture for XTE~J1751$-$305 is thus one where the powerful outburst in 2002 was associated with an episode of rapid spin-up, followed by a slow drop off in the spin-rate. One would expect this behaviour to be naturally explained in terms of a dramatic rise in the accretion torque during the initial outburst and standard magnetic dipole spin-down in between outbursts.   Assuming such a scenario will allow us to constrain the stellar magnetic field. 

Before estimating the torques required to explain the spin-evolution, we note that the evolution of XTE~J1751$-$305 is qualitatively similar to that of
another accreting pulsar, IGR~J00291+5934. 
This system underwent outbursts in 2004 and twice in rapid (34~d)
succession in 2008 (and possibly in 1998 and 2001, as well),
with each 2008 outburst being about half as luminous as the one in 2004
\cite{patruno10}.
For the 2004 outburst, the peak bolometric luminosity is
$6.3\times 10^{36}\mbox{ erg s$^{-1}$}$, assuming a distance of 5~kpc
\citep{falangaetal05},
which corresponds to $\dot{M}\sim 5.4\times 10^{-10}M_\odot\mbox{ yr$^{-1}$}$.
The 2004 outburst lasted for 14~d, during which the frequency increased with
\cite{patruno10,papittoetal11}
\begin{equation}
\dot{\nu}_{\mathrm{su}}=(5.1\pm 0.5)\times 10^{-13}\mbox{ Hz s$^{-1}$}.
 \label{eq:suigr}
\end{equation}
The first 2008 outburst had a rise-time of $<3.5\mbox{ d}$, lasted for 5~d,
and decayed with a timescale of $\approx 2\mbox{ d}$,
while the second 2008 outburst lasted for 12~d
\cite{hartmanetal11,papittoetal11}.
The 598.89~Hz pulsations associated with the spin of the neutron star had an amplitude of $\sim 10\%$ during both 2008
outbursts but was not detected in between these outbursts
(amplitude $<0.4\%$) \cite{hartmanetal11}.
The long-term decrease in frequency of pulsations detected in 2004 and
those in 2008 is $\Delta\nu=-(3-5)\times 10^{-7}\mbox{ Hz}$, yielding a
spin-down rate \cite{patruno10,hartmanetal11,papittoetal11}
\begin{equation}
\dot{\nu}_{\mathrm{sd}}=-(4\pm 2)\times 10^{-15}\mbox{ Hz s$^{-1}$}.
 \label{eq:sdigr}
\end{equation}

\begin{figure}
\begin{center}
\includegraphics[scale=0.42,clip]{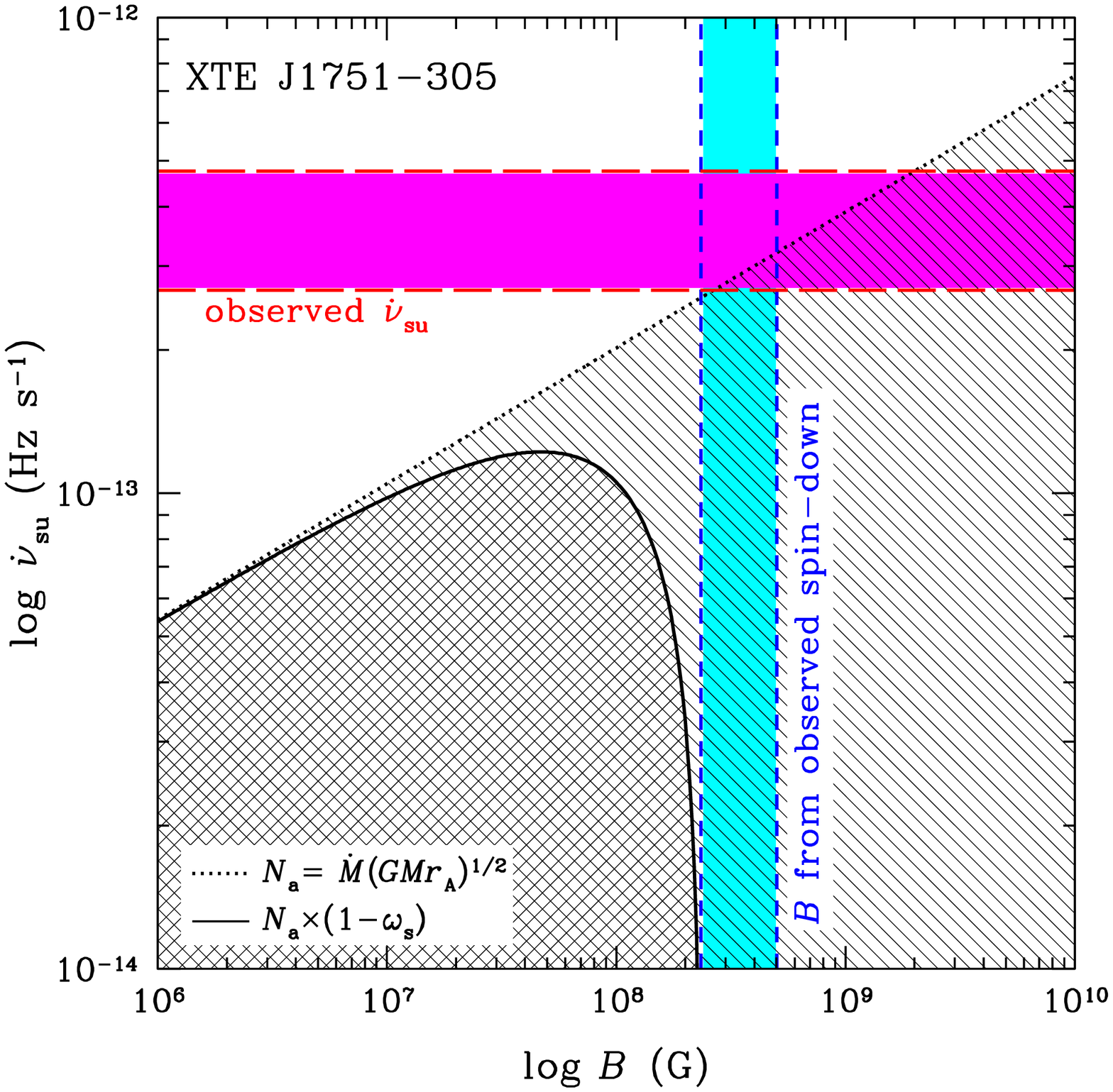}
\includegraphics[scale=0.42,clip]{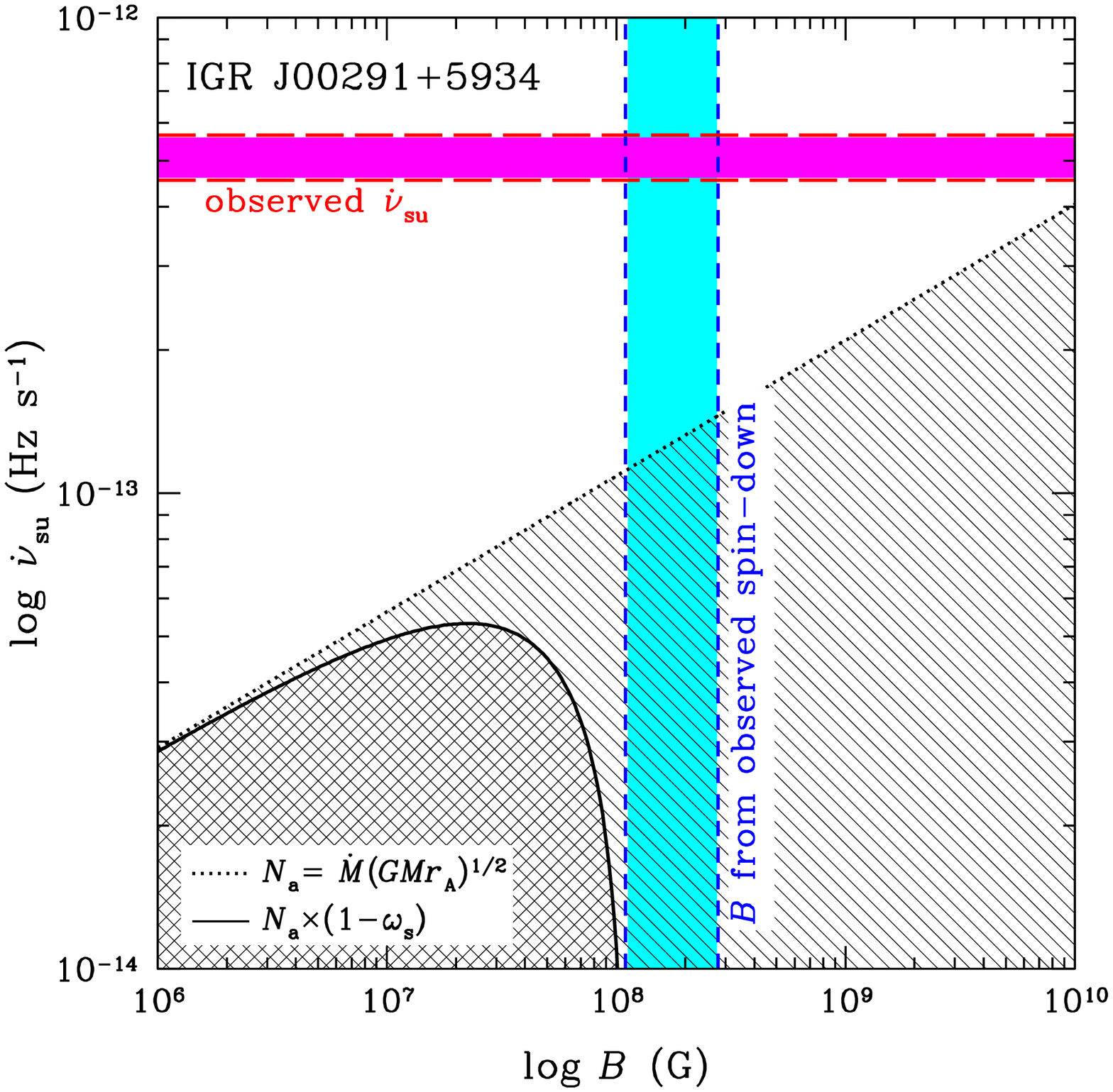}
\caption{
Magnetic field of XTE~J1751$-$305 (top) and IGR~J00291+5934 (bottom) as
determined by the observed spin-down rate, spin-up rate and fastness parameter
$\omega_{\mathrm{s}}$.
The observed spin-down rate (with 90\% uncertainty) produces the vertical band.
The \emph{observed} spin-up rate (with 90\% uncertainty) is the horizontal band.
The dotted and solid lines show the {\it maximum calculated} spin-up rates due to accretion torques
$N_{\mathrm{a}}$ and $N_{\mathrm{a}}(1-\omega_{\mathrm{s}})$, respectively.
The magnetic field determined by $\omega_{\mathrm{s}}=1$ is where the
solid line crosses the bottom axis.
}
\label{obs}
\end{center}
\end{figure}

We now consider the implications of the observed spin evolution of
XTE~J1751$-$305 and IGR~J00291+5934,
in particular using the measured changes of spin rate (or torque) to constrain
the magnetic field of each neutron star.
If the neutron star is not accreting between outbursts and the long-term
spin-down $\dot{\nu}_{\mathrm{sd}}$ is due to magnetic dipole radiation loss,
then the magnetic field at the pole is given by \citet{gunnostriker69} as
\begin{eqnarray}
B &\approx& 6.4\times 10^{19}\mbox{ G }(-\dot{\nu}_{\mathrm{sd}}/\nu^3)^{1/2}
 \nonumber\\
&=& \left\{ \begin{array}{ll}
 5.2\times 10^8\mbox{ G} & \quad\mbox{for XTE~J1751$-$305} \\
 2.8\times 10^8\mbox{ G} & \quad\mbox{for IGR~J00291+5934}
\end{array} \right. ,
\end{eqnarray}
or more accurately, by \citet{spitkovsky06} as
\begin{eqnarray}
B&\approx& 5.2\times 10^{19}\mbox{ G }[-\dot{\nu}_{\mathrm{sd}}/\nu^3
 (1+\sin^2\gamma)]^{1/2} \nonumber\\
&=& (1+\sin^2\gamma)^{-1/2} \left\{ \begin{array}{ll}
 4.3\times 10^8\mbox{ G } & \mbox{for XTE~J1751$-$305} \\
 2.2\times 10^8\mbox{ G } & \mbox{for IGR~J00291+5934}
\end{array} \right. , \label{eq:bsd}
\end{eqnarray}
where $\gamma$ is the angle between the stellar rotation and magnetic axes
(see also \citealt{contopoulosetal14}).
Figure~\ref{obs} shows, as a vertical band, the inferred magnetic field from
the observed $\dot{\nu}_{\mathrm{sd}}$ and its uncertainty
[see equations~(\ref{eq:sdxte}) and (\ref{eq:sdigr})]
and using equation~(\ref{eq:bsd}).

The horizontal band in Figure~\ref{obs} shows the \emph{observed} spin-up rate
during outburst $\dot{\nu}_{\mathrm{su}}$ and its uncertainty
[see equations~(\ref{eq:suxte}) and (\ref{eq:suigr})].
If the spin up of the neutron star (and hence the torque on the star
$2\pi I\dot{\nu}_{\mathrm{su}}$) is the result of accretion (see below),
then the magnetic field is determined by where the \emph{calculated}
spin-up due to the accretion torque intersects this horizontal band.
For example, the simple accretion torque given by
$N_{\mathrm{a}}=\dot{M}(GMr_{\mathrm{A}})^{1/2}$,
where $r_{\mathrm{A}}=(\mu^4/2GM\dot{M}^2)^{1/7}$ is the Alfv\'{e}n radius
and $\mu\sim BR^3$ is the dipole moment
(see, e.g. \citealt{pringlerees72}), is shown as the dotted line.
This torque produces a spin-up
\begin{equation}
\dot{\nu}_{\mathrm{su}}\sim 1.8\times 10^{-13}
\dot{M}_{-9}^{6/7}B_{8}^{2/7} \mbox{ Hz s$^{-1}$}\ , 
\end{equation}
where
$\dot{M}_{-9}=\dot{M}/10^{-9}M_\odot\mbox{ yr$^{-1}$}$ and
$B_{8}=B/10^{8}\mbox{ G}$.
In other words, for a measured $\dot{\nu}_{\mathrm{su}}$ and $\dot{M}$, the
neutron star magnetic field is 
\begin{eqnarray}
B &=& 1.2\times 10^7\mbox{ G }\dot{M}_{-9}^{-3}
(\dot{\nu}_{\mathrm{su}}/10^{-13}\mbox{ Hz s$^{-1}$})^{7/2} \nonumber\\
&>& \left\{ \begin{array}{ll}
 8.5\times 10^8\mbox{ G} & \quad\mbox{for XTE~J1751$-$305} \\
 2.2\times 10^{10}\mbox{ G} & \quad\mbox{for IGR~J00291+5934}
\end{array} \right. . \label{eq:bsu}
\end{eqnarray}
Here and in Figure~\ref{obs}, we use the maximal distance and peak X-ray
flux during outburst to determine $\dot{M}$. This yields the maximum torque,
and the shaded region below the dotted line denotes lower flux.
We see that the spin-up rate and spin-down rate constraints on the magnetic
field are wholly inconsistent for IGR~J00291+5934, while they are consistent,
given the observational uncertainties, for XTE~J1751$-$305.

However the simple accretion torque $N_{\mathrm{a}}$ does not fully account
for a magnetic field that threads the accretion disc and can lead to
spin-up (when $\omega_{\mathrm{s}}\lesssim 1$) or
spin-down (when $\omega_{\mathrm{s}}\gtrsim 1$),
where $\omega_{\mathrm{s}}$ $[\equiv\nu/\nu_{\mathrm{K}}(r_{\mathrm{A}})]$ is
the fastness parameter and
$\nu_{\mathrm{K}}(r_{\mathrm{A}})$ is the Kepler rotation frequency at
the Alfv\'{e}n radius.
A more realistic accretion torque is calculated by \citet{ghoshlamb79},
which is well-approximated by $N_{\mathrm{a}}(1-\omega_{\mathrm{s}})$
\citep{hoetal14}.
The spin-up regime for this torque is shown as the solid line and
cross-hatched region in Figure~\ref{obs}.
We see that this accretion torque produces a spin-up rate far below the
observed $\dot{\nu}_{\mathrm{su}}$ for XTE~J1751$-$305 and IGR~J00291+5934.
Other accretion torque estimates, such as torque
$=\dot{M}(GMr_{\mathrm{co}})^{1/2}$ or $\mu^2/r_{\mathrm{co}}^3$,
where $r_{\mathrm{co}}$ [$=(GM/\nu)^{1/3}$] is the corotation radius,
and models, such as those from \citet{wang96,kluzniakrappaport07,tauris12},
all fail to produce a strong enough spin-up torque.

Finally we can obtain a third estimate of the magnetic field by noting that
accretion occurs only when the fastness parameter $\omega_{\mathrm{s}}<1$.
This yields a magnetic field
\begin{equation}
B < 2.7\times 10^{11}\mbox{ G }\dot{M}_{-9}^{1/2}\nu^{-7/6} 
\end{equation}
That is, 
\begin{equation}
B <  \left\{ \begin{array}{ll}
 2.4\times 10^8\mbox{ G} & \quad\mbox{for XTE~J1751$-$305} \\
 1.1\times 10^8\mbox{ G} & \quad\mbox{for IGR~J00291+5934}
\end{array} \right. \ , 
\end{equation}
 indicated in Figure~\ref{obs} by where the solid line drops to zero
[since this line denotes $N_{\mathrm{a}}(1-\omega_{\mathrm{s}})$].
This estimate is close to that obtained from the dipole spin down
[see equation~(\ref{eq:bsd})], but it does not explain the spin-up rate
during outburst [see equation~(\ref{eq:bsu})]. The obvious conclusion is that
either the accretion torque is severely underestimated or the spin of the star
changes due to some other mechanism.
For example, a possible contributing factor is timing noise caused by motion
of the X-ray hot spot, such that the measured $\dot{\nu}$ is not actually a
change in the neutron star spin frequency \citep{patrunoetal09}.
However \citet{patrunoetal09} and \citet{patrunowatts12} note that timing
noise is weak in the 2002 (for XTE~J1751$-$305) and 2004 (for IGR~J00291+5934)
outburst data when $\dot{\nu}_{\mathrm{su}}$ was measured
(see equations~\ref{eq:suxte} and \ref{eq:suigr}).
Therefore it appears unlikely that timing noise could explain the entirety
of the significant discrepancy between observed $\dot{\nu}$ and calculated
$\dot{\nu}$.
Alternatively, it seems reasonable to suggest that the X-ray outburst may be associated with a crust fracture or slight rearrangement of the stellar magnetic field. This could lead to a shift in the stellar moment of inertia and a glitch similar to those seen in young radio pulsars. Noting that the observed $\Delta \nu/\nu \sim 10^{-9}$, a level typical for smaller glitches, this explanation seems a possibility. Of course, pulse-by-pulse tracking of the signal during outburst could constrain this scenario.

\section{Is there a consistent r-mode scenario?}
\label{scenarios}

So far we have arrived at two main conclusions. First of all, the value of the frequency of the observed oscillation in XTE~J1751$-$305 is consistent with an r-mode, without any particular  adjustments to our current understanding of the  physics of neutron stars, and does correspond to sensible values for mass and radius. Secondly, we have seen that, when r-modes are not included in the modelling,  the observed evolution in the spin frequency (a spin up) of the star during the X-ray outburst is not (completely) explained by the standard accretion torque. 
We will now reconsider the evolution of the spin frequency, with the r-mode included; there are a variety of possible scenarios depending on the state of the mode: stable versus unstable and unsaturated versus saturated. It is easy to anticipate that inclusion of the r-mode will make the spin evolution even more difficult to explain, not easier.  Note that \citet{strohmayermahmoodifar13} briefly described the unstable scenario and its difficulty in explaining the oscillation of XTE~J1751$-$305.

The amplitude of the mode is parameterised by $\alpha$; see \citet{oetal98} for a definition.  The results of \citet{strohmayermahmoodifar13} suggest $\alpha \sim 10^{-3}$.  One can immediately say that such a large value for an r-mode amplitude is extremely surprising on the basis of  known results.   As is discussed in detail in \citet{mahmoodifarstrohmayer13}, one can place upper limits on r-mode amplitudes of intermittently accreting neutron stars in two ways, providing one assumes the mode is active all the time (not just during outburst).  Firstly, one can use the  core temperature, inferred from the star's quiescent luminosity,  to place a limit on the rate at which an r-mode heats the star, and hence limit the mode's amplitude.  Secondly, if one assumes that the long-term time-averaged change of spin frequency, averaged over many cycles of outburst and quiescence, is zero, with a steady r-mode driven spin-down providing the balance to the (intermittent) spin-up from accretion, one again gets an upper limit on $\alpha$.  Such estimates typically lead to amplitudes in the range $10^{-8}$ to $10^{-6}$. Using the first of these procedures, \citet{mahmoodifarstrohmayer13} find $\alpha \lesssim 10^{-8}$ for XTE~J1751$-$305, much smaller than the value suggested by their measurement from the X-ray lightcurve.

However the oscillation in XTE~J1751$-$305  was observed only during the 2002 outburst itself.  One can therefore take the conservative approach of setting aside the upper limits quoted above, and consider what effect the presence of an r-mode in this system would have on the spin frequency evolution, \emph{restricting attention to the interval of the outburst}.  Two features of the observations are relevant: (i) The mode amplitude $\alpha$ has a typical value of $\sim 10^{-3}$ over the outburst duration, and (ii) the star spins up during the outburst, at an average rate given by equation (\ref{eq:suxte}) above, with a total accumulated spin up given by equation (\ref{eq:Delta_nu_observed}). We will show that, even when attention is restricted to these limited observational facts, we fail to find a self-consistent explanation for the spin evolution of the star.

The basic reason for this lies in the well known result that an r-mode of steady amplitude $\alpha$ produces a spin-down torque that would tend to make the spin frequency decrease at the rate \citep{oetal98}:
\begin{equation}
\label{eq:steady_spin_down}
\dot\Omega = \frac{2\Omega}{\tau_{\rm GW}} \alpha^2 Q ,
\end{equation}
where $Q = 9.40 \times 10^{-2}$ for a $n=1$ polytrope and $\tau_{\rm GW}$ is the timescale associated with gravitational radiation reaction effects; $\tau_{\rm GW} \approx - 1$ hour for XTE~J1751$-$305 (see equation~2.9 of \citealt{oetal98}; the negative value reflects the fact that gravitational-wave emission tends to destabilize r-modes).  
Inserting numerical values, we obtain a spin-down rate of
\begin{equation}
\dot \nu = - 2.2 \times 10^{-8} \alpha_{-3}^2 {\, \, \rm Hz \, s}^{-1} ,
\end{equation}
where $\alpha_{-3} \equiv \alpha / 10^{-3}$.  This spindown rate is more than $10^4$ times greater than the observed $\dot \nu$ over the outburst, and of opposite sign (see equation \ref{eq:suxte}).  Clearly, it is going to be difficult to reconcile the r-mode interpretation with the observed spin evolution.   Nevertheless, let us examine the different possible scenarios.

To do so we can use the evolution model of \citet{oetal98}.  This makes use of the canonical energy of the r-mode:
\begin{equation}
E_c = {1\over 2} \alpha^2 \tilde J \Omega^2 MR^2 
\label{canerg}\end{equation}
where $\tilde J=1.635\times10^{-2}$ for a $n=1$ polytrope, and also the total angular momentum of the system (star + r-mode):
\begin{equation}
J = \left( \tilde I - {3\over 2} \tilde J \alpha^2 \right) \Omega MR^2
\label{Jtot}\end{equation}
where  $\tilde I = 0.261$ (again for a $n=1$ polytrope; the constants $Q, \tilde I$ and $\tilde J$ are related by $Q = 3\tilde J / 2\tilde I$).  Together with expressions for the rate at which energy and angular momentum are radiated, these can be used to give evolution equations for the parameters $(\alpha, \Omega)$:
\begin{equation}
\dot \Omega = - 2Q \alpha^2 \Omega {1 \over \tau_\mathrm{V}} 
\label{omeq}
\end{equation}
\begin{equation}
\dot \alpha = - \alpha \left[ {1 \over \tau_\mathrm{GW} } +   {1 \over \tau_\mathrm{V}}   \right]
\label{alpeq}\end{equation}
The dimensionless amplitude $\alpha$ is expected to saturate due to non-linear effects for an unstable mode, at some value $\alpha = \alpha_{\rm sat}$.  The above pair of evolution equations are valid only for $\alpha < \alpha_{\rm sat}$. When saturation is achieved, so that $\alpha$ takes the constant value $\alpha_{\rm sat}$,  the evolution  in spin frequency is  given by
\begin{equation}
\label{eq:Omega_dot_sat}
\dot\Omega = \frac{2\Omega}{\tau_{\rm GW}} \alpha_{\rm s}^2 Q ,
\end{equation}
which is simply equation (\ref{eq:steady_spin_down}) with $\alpha  = \alpha_{\rm sat}$.  Note that the equations above neglect the (weak) spin-up accretion torque.

The current thinking is that this saturation amplitude is small, due to coupling between the r-mode and a sea of other (shorter length scale) inertial modes. The initial estimates for this effect, due to \citet{aetal03}, lead to a saturation amplitude of order $\alpha_s \approx 10^{-3}$ for a star spinning at 435~Hz, close to the value inferred from observation. 
More recent work predicts a much smaller amplitude, of $\alpha_{\rm sat} \sim 10^{-6}$ \citep{bw13}.  However, the saturation amplitude depends upon the driving mechanism, and, as we will argue below, it is difficult to reconcile an r-mode interpretation of the observations with the standard unstable r-mode scenario, so it may be that these mode saturation studies are not applicable here.

Let us now consider the different possible regimes.  The r-mode may be either stable or unstable.  Let us first consider the (more familiar) unstable case.  Before considering the evolution equations, a first obvious question, given the context provided by the  r-mode literature from the last fifteen years or so, has to be if it is conceivable that XTE~J1751$-$305 lies inside the r-mode instability window. Since this system is not the fastest or hottest among the accreting LMXBs, one would not expect this to be the case (see, e.g. Figure~2 of \citealt{Ho}). Nevertheless, we will explore the consequences of the assumption of instability.

If the mode is saturated, then equation (\ref{eq:Omega_dot_sat}) immediately predicts the rapid  spindown described above, inconsistent with the observations.  Suppose instead the mode is, at the beginning of the observation at least, not saturated.  Then equation (\ref{omeq}) shows that, for the observed value of $\alpha$, a sufficiently large value of the timescale $\tau_{\rm V}$ will give a small spin-down rate due to gravitational wave emission, opening up the possibility of a small net spin-up, once the accretion torque is included.  However, in this case equation (\ref{alpeq}) indicates that the unstable mode will grow on the short timescale $-\tau_{\rm GW} \sim 1$ hour, leading to saturation on a timescale much shorter than the observation duration of $\sim 10$ days.  Then the rapid spin-down of equation (\ref{eq:Omega_dot_sat}) again applies, in conflict with the observations.  Clearly, there is no consistent scenario in which the mode is unstable: the mode is either saturated throughout the observation, or saturates very quickly, in both cases leading to rapid spin-down, in contradiction with the observed spin-up.

Now consider the (less familiar) case of stable r-mode evolution.  In fact,  the excitation of stable r-modes and their gravitational-wave detectability has already been considered in the context of isolated pulsars  \cite{shjc12}, and an exploration of the corresponding problem in accreting millisecond pulsar systems is currently underway \citep{shjc14}.   In such scenarios, it is necessary to supply some mechanism to excite the mode in the first place, perhaps connected with a glitch in an isolated pulsar or with the process that initiates bursts in accreting systems.  We will therefore suppose some process suddenly excites the mode on some short dynamical timescale, up to an amplitude $\alpha \sim 10^{-3}$.  If this happens at fixed angular momentum, equation (\ref{Jtot}) implies a corresponding \emph{spin-up} of the star: 
\begin{equation}
\label{eq:kick_spin_up}
\Delta \Omega \sim \alpha^2 Q \Omega \Rightarrow \Delta \nu \approx 4 \times 10^{-5} \alpha_{-3}^2   {\,\, \rm Hz}.
\end{equation}
This is larger than the total spin-up measured over the course of the outburst, as given in equation (\ref{eq:Delta_nu_observed}).  However, once excited, equation (\ref{alpeq}) shows that the only way the mode can remain at (or close to) this level for the long duration of the observation $\Delta t_{\rm obs}$ is if there is a near-cancellation between the $\tau_{\rm GW}$ and $\tau_{\rm V}$ terms, so that $\tau_{\rm V} \approx -\tau_{\rm GW}$. In such a case, equation~(\ref{omeq}) again gives rapid spin-down and total spin-down over the course of the observation
\begin{equation}
\label{eq:Delta_Omega_spin_down}
\Delta \Omega = \frac{2Q\Omega \alpha^2}{\tau_{\rm GW}} \Delta t_{\rm obs}.
\end{equation}
Comparing equations (\ref{eq:kick_spin_up}) and (\ref{eq:Delta_Omega_spin_down}), we see that spin-down due to the gravitational-wave emission exceeds the initial spin-up due to the kick by a factor $\sim \Delta t_{\rm obs} / |\tau_{\rm GW}| \sim 10^2$, so that the strong gravitational-wave torque clearly results in a net spin-down over the course of the observation, in conflict with the observations.

One might attempt to evade this conclusion by, instead of exciting the mode once with an impulsive kick, having some excitation mechanism acting in a  continuous manner.  If one adds energy to $E_{\rm c}$ at a rate $\dot E_{\rm X}$, then it is straight-forward to show that the evolution equations become
\begin{equation}
\dot \Omega = - 2Q \alpha^2 \Omega \left( {1 \over \tau_\mathrm{V}} - {\dot E_\mathrm{x} \over 2E_c} \right) 
\label{omeq_X}
\end{equation}
and
\begin{equation}
\dot \alpha = - \alpha \left[ {1 \over \tau_\mathrm{GW} } +   {1 \over \tau_\mathrm{V}} - {\dot E_\mathrm{x} \over 2E_c}  \right].
\label{alpeq_X}\end{equation}
There is no difficulty in choosing a value of $\dot E_{\rm X}$ such that, in equation (\ref{alpeq_X}), $\dot \alpha$ can be made zero (or very small) such that the mode amplitude remains constant (or nearly constant) over the observation period  at  $\alpha \sim 10^{-3}$.  However, eliminating $\dot E_{\rm X}$ from (\ref{alpeq_X}) using (\ref{omeq_X}) leads to
\begin{equation}
\dot \Omega = 2Q\Omega \alpha \left[\dot\alpha + \frac{\alpha}{\tau_{\rm GW}} \right].
\end{equation}
By assumption, a nearly constant mode amplitude over the observation period means that $|\dot \alpha| \ll \alpha/|\tau_{\rm GW}|$ (this follows from $|\dot \alpha| \lesssim \alpha / \Delta t_{\rm obs}$ and $\Delta t_{\rm obs} \gg |\tau_{\rm GW}|$).  Then this equation reduces to the familiar strong spin-down of equation (\ref{eq:steady_spin_down}) above, again in conflict with the observations.

The unavoidable conclusion is that, within the standard phenomenological model for r-modes (be they unstable or stable), it is {\em not} possible to reconcile the presence of a mode excited to the suggested amplitude with the observed spin evolution of the system.

\section{Discussion}
\label{discussion}

We showed that the observed frequency of the oscillation seen in XTE~J1751$-$305 during the 2002 outburst, if interpreted as an r-mode, would correspond to sensible values of mass and radius for the neutron star.  However, we also argued that the presence of an r-mode with the large amplitude that the observations suggest, cannot be reconciled with the observed spin-up observed over the duration of the outburst.  
We conclude that the presence of an r-mode in XTE~J1751$-$305 during the 2002 outburst is not consistent with our current understanding of neutron star dynamics and must therefore be considered unlikely.

Nevertheless, one can look to  gravitational-wave observations to see if they might  be able to rule out (or indeed support) the r-mode scenario for similar events in the future. For the purpose of such a search in the case of XTE~J1751$-$305, it is interesting to note that the frequency of the signal would be known. The observed electromagnetic signal would carry the rotating frame frequency, but a gravitational-wave counterpart would be found at the inertial frame frequency. This means that one ought to search at
\begin{equation}
\nu_\mathrm{GW}  \approx \left| \kappa - m \right| \nu \approx 621\ \mathrm{Hz}.
\end{equation}
Moreover, it is easy to make a back-of-the-envelope estimate of the effective gravitational amplitude. The associated strain can be obtained from
\begin{equation}
h_0 = \sqrt{{8\pi\over 5}} {G \over c^5} {\alpha \over d} \omega_{\rm GW}^3 MR^3 \tilde J,
\end{equation}
where we take a conservative distance to the source of $d=9$~kpc and $\omega_{\rm GW} = 2\pi \nu_{\rm GW}$ \citep{owen2010}. Inserting a mode amplitude of $\alpha \sim 10^{-3}$, this leads to (for the preferred values of mass and radius from Section~\ref{mass})
\begin{equation}
h_0 \approx 1 \times 10^{-24} \alpha_{-3}.
\end{equation}
Assuming that the mode remains at this amplitude over the outburst, this would be a periodic signal, the detectability of which increases with the observation time.  The effective amplitude is then
\begin{equation}
h_\mathrm{eff} = h_0 \sqrt{ T}  
\approx 9 \times 10^{-22} \alpha_{-3} \left(\frac{\Delta T_{\rm obs}}{6 \, \rm days}\right)^{1/2}  {\,\, \rm Hz}^{-1/2} .
\end{equation}

The outburst in XTE~J1751$-$305 in fact preceded the first LIGO science run (S1) by a matter of a few months, so there is no possibility of carrying out a gravitational-wave search on archival LIGO science data.  However, we note that this would have corresponded to a signal-to-noise of about $10$ if present in LIGO S5 data, and about $100$ in Advanced LIGO data, \emph{assuming that one can perform a fully phase coherent search over the observation period}. 
 \citet{strohmayermahmoodifar13} did indeed find that the observed oscillation was highly coherent, 
quoting the oscillation frequency (in Hertz) to the sixth decimal place, i.e. at the resolution of the Fourier transform $\sim 1 /$($6$ days).  See also their Figure 4.  Clearly, the oscillation itself was intrinsically stable, and also the orbital solution was known sufficiently accurately to allow the effects of Doppler modulation due to the orbit to be largely removed over the observation interval.   Should such an accurate timing solution be available for any future outburst, a fully coherent gravitational wave search could be carried out.  Alternatively, in the absence of such an accurate timing solution, a search over a parameter space that allows for the orbital uncertainties could be carried out, but this would increase the signal-to-noise required to claim detection, as discussed in detail in \citet{wetal08}. Clearly, the signal analysis procedures for carrying out such a search merit further investigation and could be put to use if this (or a similar accreting system) were to display similar behaviour during a future science run of an advanced gravitational-wave detector.

\section*{acknowledgments}

DIJ acknowledges and thanks Ignacio (`Nacho') Santiago Prieto, Ik Siong Heng, and James Clark, with whom he has spoken about issues closely related to some of those discussed here over the last few years.  The authors thank Tod Stohmayer and Simin Mahmoodifar for feedback on an early version of the manuscript.   NA and DIJ acknowledge support from STFC via grant number ST/H002359/1, and also travel support from CompStar (a COST-funded Research Networking Programme).

\bibliographystyle{mnras}

\label{lastpage}

\end{document}